\documentclass[12pt]{iopart}
\expandafter\let\csname equation*\endcsname\relax
\expandafter\let\csname endequation*\endcsname\relax
\usepackage{amsmath}
\usepackage{iopams} 
\usepackage[english]{babel}
\usepackage{graphicx}
\usepackage{cite}
\usepackage{color}
\usepackage{bbm}
\usepackage{ulem}
\usepackage{times}
\usepackage{amsmath,amsthm,amsfonts,amssymb}
\usepackage{graphicx}
\usepackage{fancyhdr}
\usepackage{float}
\usepackage{anysize}
\usepackage{titlesec}
\usepackage{enumerate}
\usepackage{xcolor}
\usepackage{bbold}
\usepackage{braket}

\usepackage{tikz,xcolor,hyperref}

\definecolor{lime}{HTML}{A6CE39}
\DeclareRobustCommand{\orcidicon}{%
	\begin{tikzpicture}
	\draw[lime, fill=lime] (0,0) 
	circle [radius=0.16] 
	node[white] {{\fontfamily{qag}\selectfont \tiny ID}};
	\draw[white, fill=white] (-0.0625,0.095) 
	circle [radius=0.007];
	\end{tikzpicture}
	\hspace{-2mm}
}

\foreach \x in {A, ..., Z}{%
	\expandafter\xdef\csname orcid\x\endcsname{\noexpand\href{https://orcid.org/\csname orcidauthor\x\endcsname}{\noexpand\orcidicon}}
}

\begin{document}

\title{Steering spin fluctuations in lattice systems via two-tone Floquet engineering} 

\author{Ruben Pe\~{n}a \orcidA{}}
\address{Departamento de F\'isica, Universidad de Santiago de Chile, 
Avenida V\'ictor Jara 3493, 9170124, Santiago, Chile}
\address{BCAM - Basque Center for Applied Mathematics, Mazarredo, 14 E48009 Bilbao, Basque Country – Spain}

\author{Felipe Torres \orcidB{}}
\address{Departamento de F\'isica, Facultad de Ciencias,
Universidad de Chile, Casilla 653, Santiago, Chile 7800024}
\address{Center for the Development of Nanoscience and Nanotechnology, Estaci\'on Central, 9170124, Santiago, Chile}

\author{Guillermo Romero\orcidC{}}
\address{Departamento de F\'isica, CEDENNA, Universidad de Santiago de Chile, 
Avenida V\'ictor Jara 3493, 9170124, Santiago, Chile}
\ead{guillermo.romero@usach.cl}
\date{\today}

\begin{abstract}
We report on the control of spin pair fluctuations using two-tone Floquet engineering.  We consider a one-dimensional spin-1/2 lattice with periodically modulated spin exchanges using parametric resonances. The stroboscopic dynamics generated from distributed spin exchange modulations lead to spin pair fluctuations reaching quasi-maximally correlated states and a subharmonic response in local observables, breaking the discrete-time translational symmetry. We present a protocol to control the interacting many-body dynamics, producing spatial and temporal localization of correlated spin pairs via dynamically breaking correlated spin pairs from the edges towards the center of the lattice. Our result reveals how spin fluctuations distribute in a heterogeneous lattice depending on parametric resonances. This may open new routes for exploring distinct nonequilibrium states of matter and the conduction of quasiparticles in quantum materials.
\end{abstract}
\maketitle

\section{Introduction}
Controlling nonequilibrium dynamics of interacting many-body systems poses significant challenges because of the spreading of entanglement; however, it is an area of active research due to its potential applications in quantum technologies, including quantum metrology \cite{D.Lukin} and sensing \cite{montenegro2024}. Floquet engineering relies on the control of nonequilibrium quantum systems by using periodic drivings \cite{RevModPhys.89.011004,doi:10.1146/annurev-conmatphys-031218-013423,Bukov2015Mar,PhysRevX.4.031027}. The advent of Floquet engineering has been motivated by theoretical predictions on nonequilibrium states of matter without static analog such as discrete time crystals \cite{PhysRevA.91.033617,PhysRevLett.117.090402,PhysRevLett.118.030401,Sacha_2017,PhysRevLett.123.150601,TimeCrystals,Pizzi:2021we,PhysRevLett.127.140602,PhysRevB.104.094308,PhysRevResearch.3.L042023,PhysRevResearch.5.023014}, dynamical many-body freezing \cite{PhysRevB.82.172402,PhysRevB.90.174407}, Floquet dynamical phase transitions \cite{PhysRevB.106.094314}, and Floquet prethermal states \cite{Abanin:2017uy,PhysRevB.95.014112}. Nowadays, the development of state-of-the-art quantum simulators \cite{Buluta2009Oct,Ippoliti2021Sep,PRXQuantum.2.017003} has pushed the interest forward since they allow us to stabilize nonequilibrium states of matter, including prethermal discrete time-crystals \cite{Kyprianidis2021Jun,Beatrez2023Mar}, time-crystalline eigenstate order \cite{Mi2022Jan}, and Floquet prethermal states \cite{PhysRevX.10.021044,PhysRevA.105.012418}. 

Periodic driving of quantum systems leads to dressed states described by the time-independent Floquet Hamiltonian $\hat{H}_F$. When the driving frequency is larger than any frequency scale of the undriven quantum system (high-frequency regime), $\hat{H}_F$ can be defined approximately using the Magnus expansion \cite{Magnus1,Blanes_2010}. The stroboscopic nonequilibrium dynamics of the system are described by the effective Floquet Hamiltonian \cite{RevModPhys.89.011004,doi:10.1146/annurev-conmatphys-031218-013423,Bukov2015Mar,PhysRevX.4.031027}.

In recent contributions, it has been pointed the emergence of many-body resonances in the high-frequency regime \cite{Bukov2016Apr,DallaTorre2021Aug,Pena2022Aug,Pena2022Dec,Li2022Oct,Ha2023Jun}. In particular, Refs. \cite{Pena2022Aug,Pena2022Dec} consider many-body parametric resonances that may activate nearest-neighbor or next-nearest-neighbor interactions depending on what parametric resonance is chosen. Those many-body parametric resonances appear in a broad class of many-body Hamiltonians exhibiting U(1) and parity symmetry such as the Bose-Hubbard model \cite{PhysRevB.40.546,Jaksch1998}, the $XXZ$ spin-1 model \cite{PhysRevB.67.104401,PhysRevLett.126.163203} or the Jaynes-Cummings-Hubbard model \cite{Greentree:2006aa,Hartmann:2006aa,PhysRevA.76.031805}. Also, recently focusing on two-tone driving protocols has opened up possibilities for further extending Floquet engineering \cite{Beatrez2023Mar,PhysRevA.107.043309,PhysRevResearch.4.013056}. Two-tone Floquet engineering refers to specific protocols where the quantum system is subjected to a periodic driving characterized by two different frequencies. Since parametric resonances exist in many-body quantum systems \cite{Bukov2016Apr,DallaTorre2021Aug,Pena2022Aug,Pena2022Dec,Li2022Oct,Ha2023Jun}, natural questions arise: how do the parametric resonances influence the dynamics of many-body systems under a two-tone periodic protocol? what novel nonequilibrium situation may arise when the applied driving frequencies are modified? 

This work addresses the previous questions and provides crucial insights. We report on controlling the nonequilibrium dynamics of an interacting spin system, producing spatial and temporal localization of quasi-maximally correlated spin pairs using two-tone Floquet protocols. We consider a one-dimensional spin-1/2 lattice with periodically modulated spin exchanges using parametric resonances acting upon consecutive spin-spin exchange. This two-tone Floquet engineering leads to a subharmonic response in local observables, thus spontaneously breaking the discrete-time translational symmetry and enabling the control of spin fluctuations along the lattice. The control of these spin fluctuations is reached using two different two-tone Floquet protocols leading to one-period evolution operators $U_1(T)$ and $U_2(T)$, where $T=2\pi/\Omega_0$, which are consecutively applied to the spins system. To demonstrate the control of the spin pair fluctuations in the spin-1/2 lattice, we use the Magnus expansion to derive the effective Hamiltonian that captures the dominant spin fluctuations governing the system dynamics. 

This article is organized as follows. In section \ref{MBR}, we identify the emergence of many-body resonances in the transverse field Ising model with open boundary conditions and a modulated spin exchange. We present numerical simulations of
fidelity susceptibility to identify many-body resonances. In section \ref{FE}, we use Floquet theory to analyze the emergent many-body dynamics in a spins lattice with periodically modulated spin exchanges using the fundamental frequency and its first harmonic acting upon consecutive spin exchanges. We start our discussion with the three-spin lattice to prove the control of spin pair fluctuations when one spin exchange is driven with the first harmonic, whereas the second spin exchange is driven with the fundamental frequency. Then, we extend our investigation to the many-body case. We present numerical simulations of non-local observables, such as nearest-neighbor correlation functions, to demonstrate the control of spin pair fluctuations in a lattice of size $L$. In section \ref{sec:4}, we propose a novel Floquet protocol to control the interacting many-body dynamics, producing spatial and temporal localization of quasi-maximally correlated spin pairs. This is achieved via dynamically breaking correlated spin pairs from the edges towards the center of the lattice in the one-dimensional spin-1/2 lattice. The latter is reached using two different two-tone Floquet protocols leading to one-period evolution operators $\hat{U}_1(T)$ and $\hat{U}_2(T)$, which are consecutively applied to the spins system. Finally, in section \ref{Cl}, we present our concluding remarks.

\section{Many-body resonances in the transverse field Ising model}
\label{MBR}
The transverse field Ising model (TFIM) is a ubiquitous Floquet-engineered spin system that describes $L$ interacting spin-$1/2$ particles. Considering open boundary conditions and a modulated spin exchange, the time-dependent Hamiltonian reads 
\begin{equation}
\hat{H}_{\rm TFIM}(t)=\hat{H}_{0} +\hbar  J_0\cos{(\Omega t)}\sum_{j=1}^{L-1}\hat{\sigma}^{x}_j\hat{\sigma}^{x}_{j+1},
\label{TFIM}
\end{equation}
where $\hat{H}_{0}=\hbar g\sum^L_{j=1}\hat{\sigma}_j^z$ represents the local energy term. The transverse magnetic field, the bare exchange coupling, and the driving frequency are denoted by $g$, $J_0$, and $\Omega$, respectively. We use this model as proof of concept but also because of the feasibility of implementing this model in trapped ions \cite{Kim2011Oct} and optical lattices \cite{Pieplow_2018}. The operators $\hat{\sigma}_j^{\alpha}$ with $\alpha=x,z$, denote the Pauli matrices at lattice site $j$. The operator $\hat{\sigma}_j^z$ satisfies the eigenvalue equation $\hat{\sigma}^z_j\ket{m_j}=m_j\ket{m_j}$, where $m_j$ takes values $\{-1,1\}$, and it represents the $z$ component of the spin. Along with this work, we consider the regime $g/J_0 \gg 1$, where the local magnetic field dominates over the spin-spin exchange. This regime will permit the emergence of many-body resonances \cite{Bukov2016Apr,DallaTorre2021Aug,Pena2022Aug,Pena2022Dec,Li2022Oct,Ha2023Jun} and quasi-maximally correlated spin pairs in the spin-lattice system. 

In what follows, we briefly explain the emergence of many-body resonances in the system described by the Hamiltonian (\ref{TFIM}). Setting the initial state of the system as the product state $\ket{\psi_0}=\bigotimes_{j=1}^{L}\ket{\downarrow_j}$. A spin exchange event will flip nearest-neighbor spins $j$ and $j+1$ leading to several configurations span by the set $\mathcal{S}_1=\{\ket{\downarrow_1,...,\uparrow_j,\uparrow_{j+1},...,\downarrow_L}\}_{j=1}^{L-1}$. To identify one possible many-body resonance, we compute the free energy of the initial configuration $\ket{\psi_0}$ and one possible state within the set $\mathcal{S}_1$ by considering the Hamiltonian $\hat{H}_0$, that is,
\begin{eqnarray}
\hat{H}_{0}\ket{\downarrow_1,...,\downarrow_j,\downarrow_{j+1},...,\downarrow_L}&=-Lg\hbar\ket{\downarrow_1,...,\downarrow_j,\downarrow_{j+1},...,\downarrow_L},\nonumber\\
\hat{H}_{0}\ket{\downarrow_1,...,\uparrow_j,\uparrow_{j+1},...,\downarrow_L}&=4g\hbar -Lg\hbar\ket{\downarrow_1,...,\uparrow_j,\uparrow_{j+1},...,\downarrow_L}.
\label{Hlocal}
\end{eqnarray}
Notice that any possible configuration within $\mathcal{S}_1$ has local energy $4g\hbar -Lg\hbar$. 
The energy difference between the initial state and any configuration within $\mathcal{S}_1$ is $\Delta E= 4g\hbar$. Therefore, to activate the nearest-neighbor spin exchange, the driving frequency should match the condition $\Omega_1=4g$. This is analog to many-body resonances in strongly interacting boson systems\cite{DallaTorre2021Aug,Pena2022Aug,Pena2022Dec}. As has been proved in Refs.\cite{Pena2022Aug,Pena2022Dec}, many-body resonances may activate long-range interactions as next-nearest-neighbor (NNN) interactions depending on the driving frequency. In the modulated TFIM (\ref{TFIM}), NNN spin exchange will occur via a second-order process with the driving frequency satisfying the condition $\Omega_0 = 2g$. In this case, the initial configuration $\ket{\downarrow_1,...,\downarrow_j,\downarrow_{k},\downarrow_l,...,\downarrow_L}$ will partially exchange population with states within the set $\mathcal{S}_2=\{\ket{\downarrow_1,...,\uparrow_j,\downarrow_{k},\uparrow_l,...,\downarrow_L}\}$ via two spin-exchange events that flip the spins $j$ and $l$. This picture of many-body resonances is valid within the regime $g/J_0\gg 1$ \cite{Pena2022Aug,Pena2022Dec}.

\begin{figure}[t]
\centering
\includegraphics[scale=0.4]{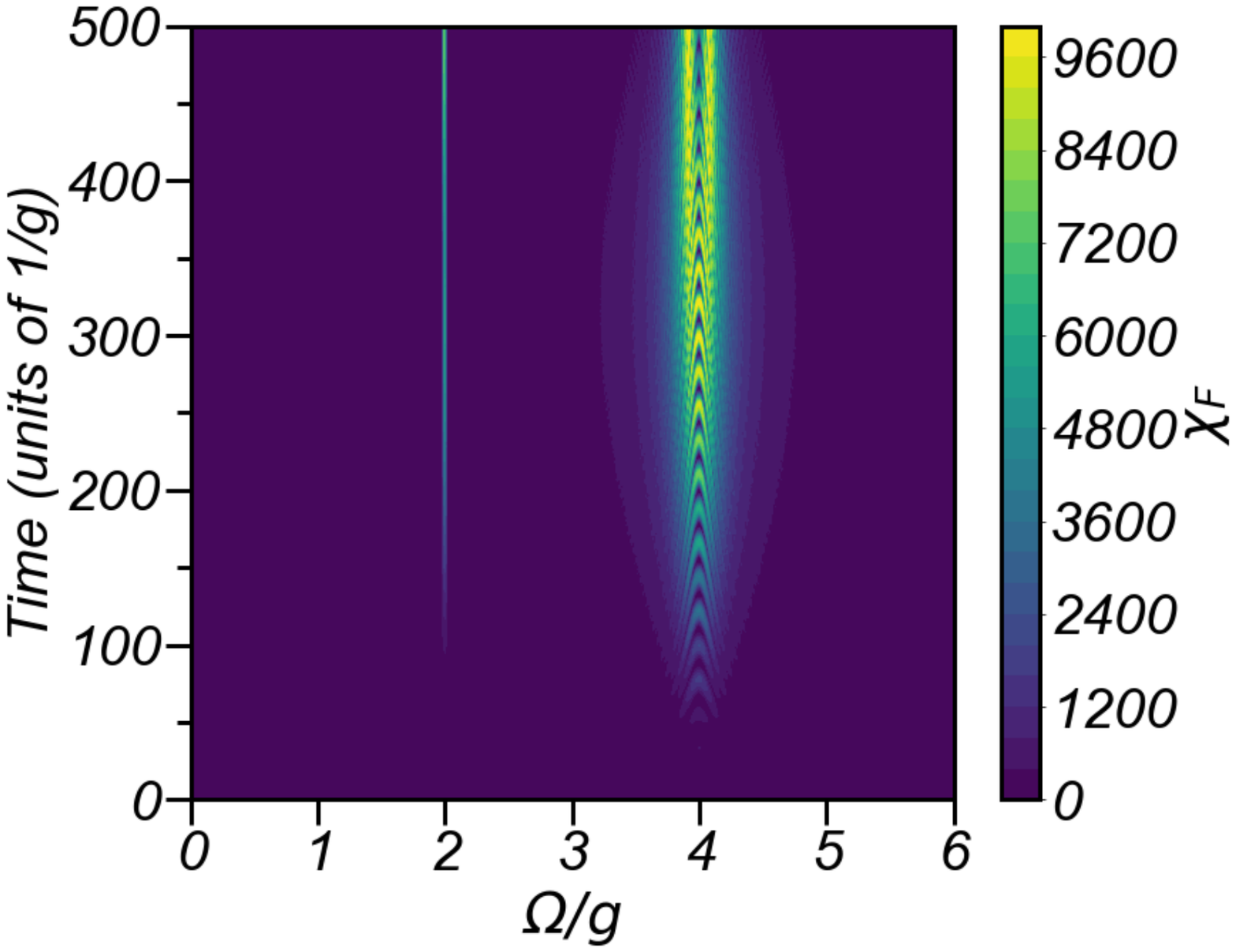}
\caption{Fidelity susceptibility $\chi_F(\Omega,t)$ as a function of the driving frequency $\Omega$ and time $t$. We consider a spin-lattice of size $L=3$ described by the Hamiltonian (\ref{TFIM}). Details of the numerical calculation appear in the main text.}
\label{Fig1}
\end{figure}
Another way to recognize many-body resonances in the spin-lattice system is to use the whole quantum many-body state and fidelity susceptibility (FS). The latter quantifies the fidelity response to a slight change of driving parameter acting upon the quantum system \cite{You2007Aug,Gu2008Jun,Gu2010Sep}. In our case, the FS will be a function of time $t$ and the driving frequency $\Omega$, and is defined as $\chi_F(\Omega,t)=\langle \partial_{\Omega}\psi(\Omega,t)\vert \partial_{\Omega}\psi(\Omega,t)\rangle - \vert \langle\partial_{\Omega}\psi(\Omega,t)\vert \psi(\Omega,t)\rangle \vert^2$. In Fig.\ref{Fig1}, we plot $\chi_F(\Omega,t)$ for a $L=3$ spin-lattice described by the Hamiltonian (\ref{TFIM}). The quantum state $\ket{\psi(\Omega,t)}$ is numerically calculated using the evolution operator for a time-dependent Hamiltonian, with initial state $\ket{\psi_0}=\bigotimes_{j=1}^{3}\ket{\downarrow_j}$. All parameters are defined in terms of the local transverse magnetic field $g$, that is, $J_0=0.1g$, the driving frequency $\Omega \in [0.0g,6.0g]$ with a frequency step size of $\delta\Omega=0.01g$, and time step $\delta t=0.01g^{-1}$.  Figure \ref{Fig1} shows that $\chi_F(\Omega,t)$ exhibits abrupt changes near the fundamental frequency $\Omega_0$ and the first harmonic $\Omega_1$. This way, using FS and the whole quantum many-body state allows us to predict the emergence of parametric many-body resonances. We stress that parametric resonances predicted in this model may be achievable in superconducting devices where driving microwave frequencies may range $\Omega/2\pi\sim 0-18$ GHz and $g/2\pi\sim 4-8$ GHz \cite{Forn-Diaz2010,Chang2020Jan}.
 
Our previous results consider a spin-lattice with equally modulated spin exchanges. Now, we will focus on two-tone drivings in the spins system using parametric resonances. Surprisingly, modulating the lattice inhomogeneously within Floquet engineering will lead to spin pair fluctuations reaching quasi-maximally correlated states and the spontaneous breaking of the discrete-time translational symmetry evidenced in local observables. These dynamic features of our two-tone Floquet protocols will allow us to control and distribute spin pair fluctuations along the lattice.
\begin{figure}[t]
\centering
\includegraphics[scale=0.5]{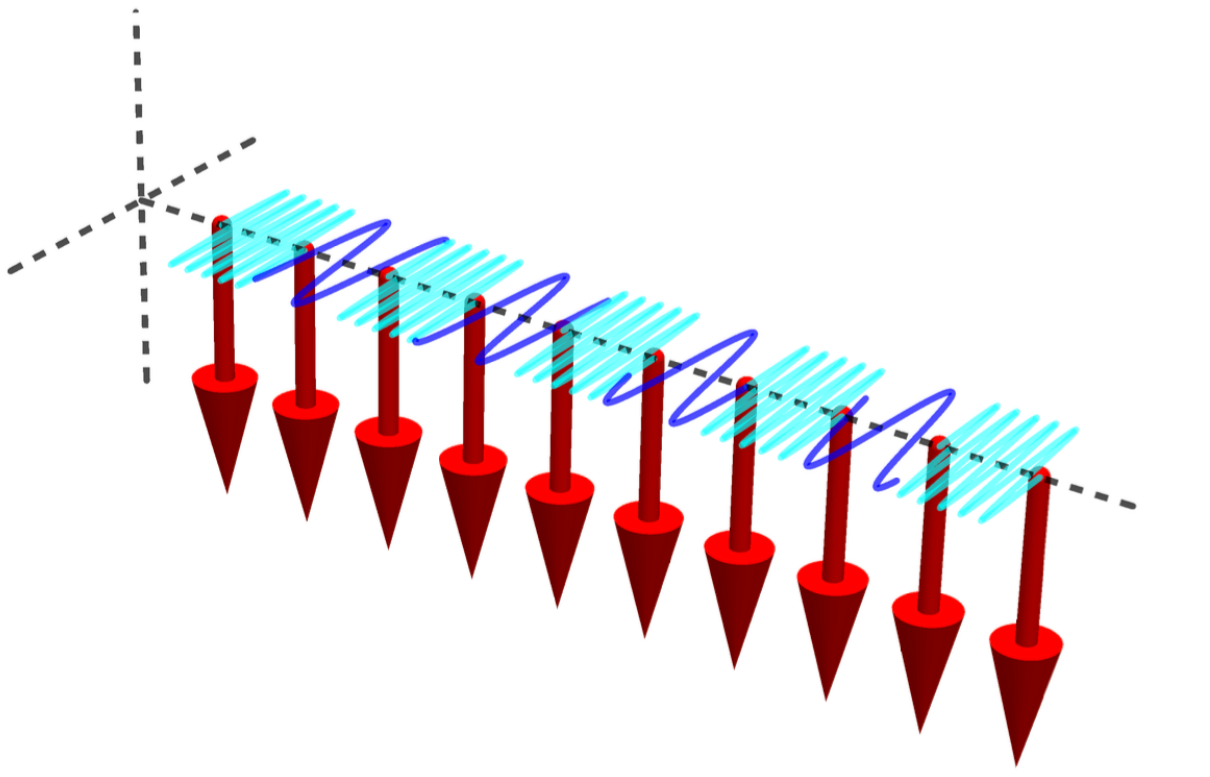}
\caption{Schematic representation of nearest-neighbor interacting spins with periodically modulated exchange rates. The wiggle blue curve represents the fundamental frequency $\Omega_0$, whereas the light-blue curve represents the frequency $\Omega_1=2\Omega_0$.}
\label{Fig2}
\end{figure}

\section{Two-tone Floquet engineering}
\label{FE}
This section addresses a spins lattice with inhomogeneous modulated spin exchanges; see figure~\ref{Fig2} for a schematic representation. We will prove later using fidelity susceptibility \cite{You2007Aug,Gu2008Jun,Gu2010Sep} the emergence of integer and fractional parametric resonances. Integer resonances will allow us to define our two-tone Floquet protocols. We start our discussion with the three-spin lattice in what follows, representing the minimal setup required to control spin pair fluctuations in the lattice and subharmonic response in local observables. 

\subsection{Three-spin lattice}
\begin{figure}[t]
\centering
\includegraphics[scale=0.4]{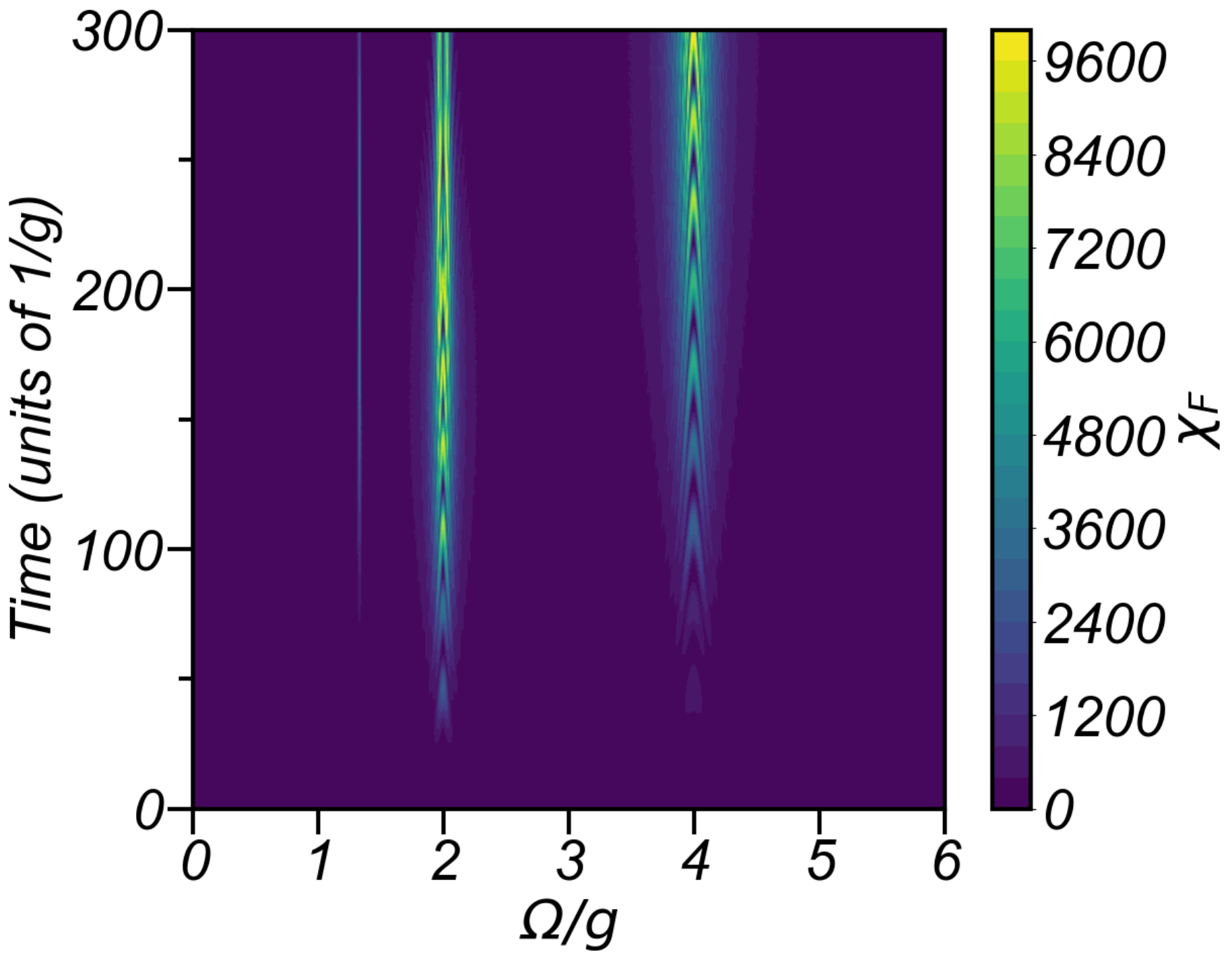}
\caption{Fidelity susceptibility $\chi_F(\Omega,t)$ as a function of the driving frequency $\Omega$ and time $t$. We consider a spin-lattice of size $L=3$ described by the Hamiltonian (\ref{TrimerTwo}). All parameters are defined in terms of the local transverse magnetic field $g$, that is, $J_0=0.1g$ and the driving frequency $\Omega \in [0.0g,6.0g]$ with a frequency step size of $\delta\Omega=0.01g$.}
\label{Fig3}
\end{figure}
Let us consider a three-spin lattice with interleaved drivings where one spin exchange is driven with frequency $2\Omega$, whereas the second spin exchange is driven with frequency $\Omega$.  The initial condition is the product state $\ket{\psi_0}=\ket{\downarrow \downarrow \downarrow}$. The three-spin lattice Hamiltonian reads
\begin{equation}
\hat{H}(t)=\hbar g\sum^3_{j=1}\hat{\sigma}_j^z +  \hbar J_0\cos{(2\Omega t)}\hat{\sigma}^{x}_1\hat{\sigma}^{x}_{2} + \hbar J_0\cos{(\Omega t)}\hat{\sigma}^{x}_2\hat{\sigma}^{x}_{3}.
\label{TrimerTwo} 
\end{equation}

We numerically solved the Sch\"odinger equation using the Hamiltonian (\ref{TrimerTwo}) and computed fidelity susceptibility to recognize parametric resonances. Figure \ref{Fig3} shows that fidelity susceptibility $\chi_F(\Omega,t)$ exhibits abrupt changes near the frequencies $\Omega_0$, $\Omega_1$, and a thinner line at $\Omega_2=(4/3)g$. Integer resonances $\Omega_0$ and $\Omega_1$ also appear in the homogeneous driving case (c.f. figure \ref{Fig1}). The fractional resonance $\Omega_2$ will lead to second-order processes activating next-nearest-neighbor spin exchange under the condition $\Omega=\Omega_2$, similar to the case already studied in systems with $U(1)$ symmetry \cite{Pena2022Aug,Pena2022Dec}. As we will prove next, the simultaneous application of parametric resonances $\Omega_0$ and $\Omega_1$ will lead to spin pair fluctuations in the lattice.  

Now, let us consider the scenario where the driving frequency in the Hamiltonian \ref{TrimerTwo} is set to  $\Omega=\Omega_0$ and we consider the case $g/J_0\gg 1$. The latter implies we are working in the high-frequency regime of modulated hopping rates \cite{Bukov2015Mar}. The initial condition is the product state $\ket{\psi_0}=\ket{\downarrow \downarrow \downarrow}$.
Since $\Omega_1$ is twice the fundamental frequency $\Omega_0$, the Hamiltonian (\ref{TrimerTwo}) is periodic $\hat{H}(t)=\hat{H}(t+T)$ with period $T=2\pi/\Omega_0$. Therefore, we can apply the Floquet theory to time-periodic Hamiltonians. In Floquet theory \cite{Floquet1883,Grifoni1998Oct,Bukov2015Mar}, the main features of the system dynamics can be captured by the one-period evolution operator $\hat{U}(T+t_0,t_0)=e^{-i\hat{H}_F[t_0]T/\hbar}$, where $\hat{H}_F[t_0]$ is the time-independent Floquet Hamiltonian. Finding $\hat{H}_F[t_0]$ is challenging in a generic situation. Still, if the frequency driving is larger than any frequency scale of the undriven system, the Floquet Hamiltonian can be defined approximately using the Magnus expansion \cite{Bukov2015Mar}. The first term reads as
\begin{align}
\label{HF01}
\hat{H}^{(0)}_F= \frac{1}{T}\int_0^Tdt \hat{H}(t). 
\end{align}

Since the transverse field $g$ is large as compared to $J_0$, it is convenient to move to a rotating frame considering the free Hamiltonian $\hat{H}_0=\hbar g\sum_{j=1}^3\hat{\sigma}^z_j$. We obtain
\begin{align}
\hat{H}_I(t)&= \hbar J_0\cos{(\Omega_1t)}\big(e^{4igt}\hat{\sigma}^{+}_1\hat{\sigma}^{+}_{2} +e^{-4igt} \hat{\sigma}^{-}_1\hat{\sigma}^{-}_{2}+ \hat{\sigma}^{+}_1\hat{\sigma}^{-}_{2}+ \hat{\sigma}^{-}_1\hat{\sigma}^{+}_{2}\big)\nonumber\\
&+ \hbar J_0\cos{(\Omega_0t)}\big(e^{4igt}\hat{\sigma}^{+}_2\hat{\sigma}^{+}_{3} +e^{-4igt} \hat{\sigma}^{-}_2\hat{\sigma}^{-}_{3}+ \hat{\sigma}^{+}_2\hat{\sigma}^{-}_{3}+ \hat{\sigma}^{-}_2\hat{\sigma}^{+}_{3}\big).
\label{TrimerHI}
\end{align}

We compute the Floquet Hamiltonian by replacing the Hamiltonian (\ref{TrimerHI}) in the Eq.~(\ref{HF01}). In this case, we obtain 
\begin{align}
\hat{H}^{(0)}_F=\frac{\hbar J_0}{2}\big(\hat{\sigma}^{+}_1\hat{\sigma}^{+}_{2} + \hat{\sigma}^{-}_1\hat{\sigma}^{-}_{2}\big).
\label{HF02}  
\end{align}

The three-spin system governed by the Hamiltonian (\ref{HF02}) will only access states $\ket{\psi_0}=\ket{\downarrow \downarrow \downarrow}$ and $\ket{\psi_1}=\ket{\uparrow \uparrow \downarrow}$ along the dynamics. Therefore, the Sch\"odinger equation can be solved analytically by diagonalizing the  Hamiltonian represented by a 2 × 2 matrix within the effective basis formed by states $\{\ket{\psi_0},\ket{\psi_1}\}$, that is
\begin{equation}
\hat{H}^{(0)}_F=
\hbar \left(
\begin{array}{cc}
0&\frac{J_0}{2}\\
\frac{J_0}{2}&0\\
\end{array}
\right).
\label{HF03}
\end{equation}

The wave function at time $t$ reads
\begin{equation}
\ket{\psi(t)}=\cos{\bigg(\frac{J_0}{2}t\bigg)}\ket{\psi_0} - i\sin{\bigg(\frac{J_0}{2}t\bigg)}\ket{\psi_1}.
\end{equation}
The analytical expectation values of operators $\hat{\sigma}^z_j$ ($j=1,2,3$) at stroboscopic times $t=nT$ read
\begin{align}
\langle \hat{\sigma}^z_3(nT) \rangle &=-1,
\end{align}
\begin{align}
\langle \hat{\sigma}^z_1(nT) \rangle=\langle \hat{\sigma}^z_2(nT) \rangle&=-\cos{\bigg(\frac{2 \pi J_0}{\Omega_0} n \bigg)}.
\label{Ana}
\end{align}

\begin{figure}[t]
\centering
\includegraphics[scale=0.65]{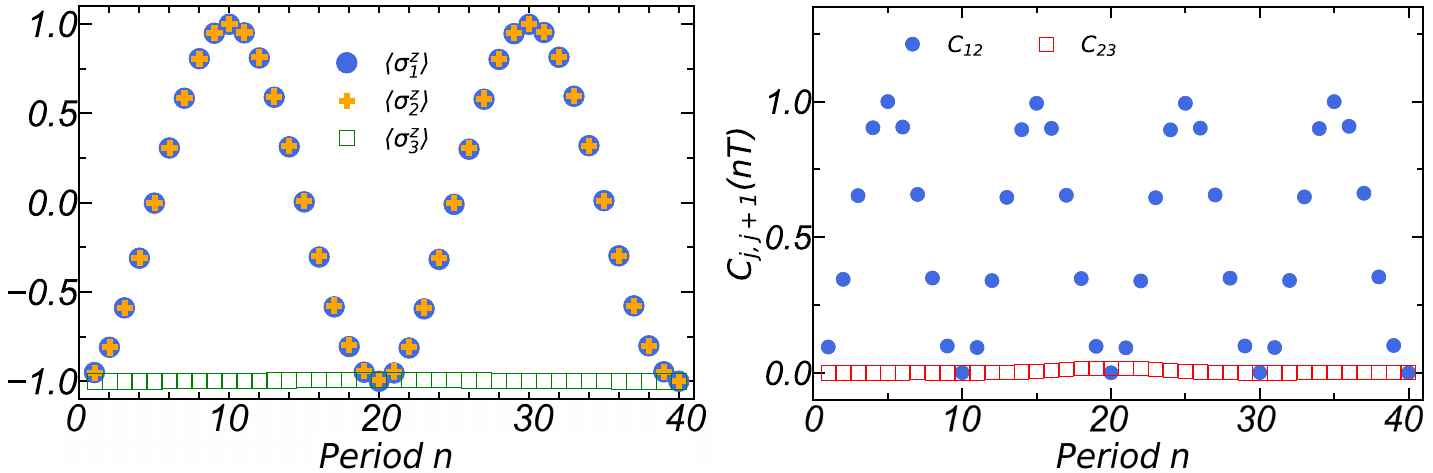}
\caption{The left panel shows expectation values of $\hat{\sigma}^z_j$ at each lattice site $j$, while the right panel shows nearest-neighbor correlation functions. The figure shows the three-spin lattice where the first spin exchange is driven with frequency $\Omega_1$, whereas the second spin exchange is driven with frequency $\Omega_0$. The three-spin lattice is initialized in the product state $\ket{\psi_0}=\bigotimes_{j=1}^{L}\ket{\downarrow_j}$, and we use the parameter $J_0=0.1g$ and time step $\delta t=0.001g^{-1}$.}
\label{Fig4}
\end{figure}
In the left panel of Fig \ref{Fig4}, we show the expectation values of $\hat{\sigma}^z_j$ for each lattice site at stroboscopic times. We computed the one-period evolution operator $\hat{U}(T)$ in this numerical simulation using the Hamiltonian (\ref{TrimerTwo}), with the driving frequency set to $\Omega=\Omega_0$. As a result, the quantum state at stroboscopic times read $\ket{\psi(nT)}=\hat{U}(T)\ket{\psi_0}$, where $T=2\pi/\Omega_0$. It can be seen that after ten periods, a spin exchange flips spins $1$ and $2$ from the state $\ket{\psi_0}$ to state $\ket{\psi_1}$. The analytical results in Eq.~(\ref{Ana}) show that the periodicity of the expectation value of the operators $\hat{\sigma}^z_1$ and $\hat{\sigma}^z_2$ is $T'=2 \pi/J_0$, which satisfies the relation $T'=20T$. Therefore, the system has a subharmonic response, leading to the spontaneous breaking of the discrete-time translational symmetry \cite{TimeCrystals}. Also, the rightmost spin is blocked, since $\langle\hat{\sigma}^z_3(nT)\rangle$ remains static.
In the right panel of Fig \ref{Fig4}, we show the nearest-neighbour correlation functions defined as $C_{j,j+1}= \langle \hat{\sigma}^z_j \hat{\sigma}^z_{j+1} \rangle - \langle \hat{\sigma}^z_j \rangle \langle \hat{\sigma}^z_{j+1} \rangle$ at stroboscopic times. The spin pair fluctuations between the first two spins achieve quasi-maximally correlated states, reaching a value of $0.99$ at $5T$. Due to the symmetry with respect to the center spin, spin pair fluctuations can be generated between the center and rightmost spins if the driving frequency of the first spin exchange is equal to $\Omega_0$ and of the second spin exchange is $\Omega_1$. As discussed in the next section, these dynamic features of our two-tone Floquet protocols will be critical for steering spin fluctuations along the lattice.

The previous results consider a three-spin lattice with inhomogeneous modulated spin exchanges, and we identify spin pair fluctuations in the spins system. The next section will extend our investigation to the many-body case. In particular, we will prove that spin pair fluctuations emerge in a lattice of size $L$ with open boundary conditions.

\subsection{Many-body lattice}
This section considers a many-body lattice of size $L$ with interleaved drivings involving parametric resonances $\Omega_0$ and $\Omega_1$ acting upon consecutive spin exchanges. The Hamiltonian reads
\begin{align}
\hat{H}(t)&=\hbar g\sum^L_{j}\hat{\sigma}_j^z +  \hbar J_0\cos{(\Omega_1 t)}\sum^{L-1}_{j~\rm odd}\hat{\sigma}^{x}_j\hat{\sigma}^{x}_{j+1} + \hbar J_0\cos{(\Omega_0 t)}\sum^{L-1}_{j~\rm even}\hat{\sigma}^{x}_j\hat{\sigma}^{x}_{j+1}.
\label{MB} 
\end{align}
We move to a rotating frame taking $\hat{H}_0=\hbar g\sum^L_{j}\hat{\sigma}_j^z $. The Hamiltonian reads 
\begin{eqnarray}
\hat{H}_I(t)&=\hbar J_0\cos{(\Omega_1 t)}\sum^{L-1}_{ j~\rm odd}\big(e^{4igt}\hat{\sigma}^{+}_j\hat{\sigma}^{+}_{j+1} +e^{-4igt} \hat{\sigma}^{-}_j\hat{\sigma}^{-}_{j+1}+ \hat{\sigma}^{+}_j\hat{\sigma}^{-}_{j+1}+ \hat{\sigma}^{-}_j\hat{\sigma}^{+}_{j+1}\big)\nonumber\\ 
&+\hbar J_0\cos{(\Omega_0 t)}\sum^{L-1}_{j~\rm even}\big(e^{4igt}\hat{\sigma}^{+}_j\hat{\sigma}^{+}_{j+1} +e^{-4igt} \hat{\sigma}^{-}_j\hat{\sigma}^{-}_{j+1}+ \hat{\sigma}^{+}_j\hat{\sigma}^{-}_{j+1}+ \hat{\sigma}^{-}_j\hat{\sigma}^{+}_{j+1}\big).
\label{MBHI}
\end{eqnarray}

\begin{figure}[t]
\centering
\includegraphics[scale=0.32]{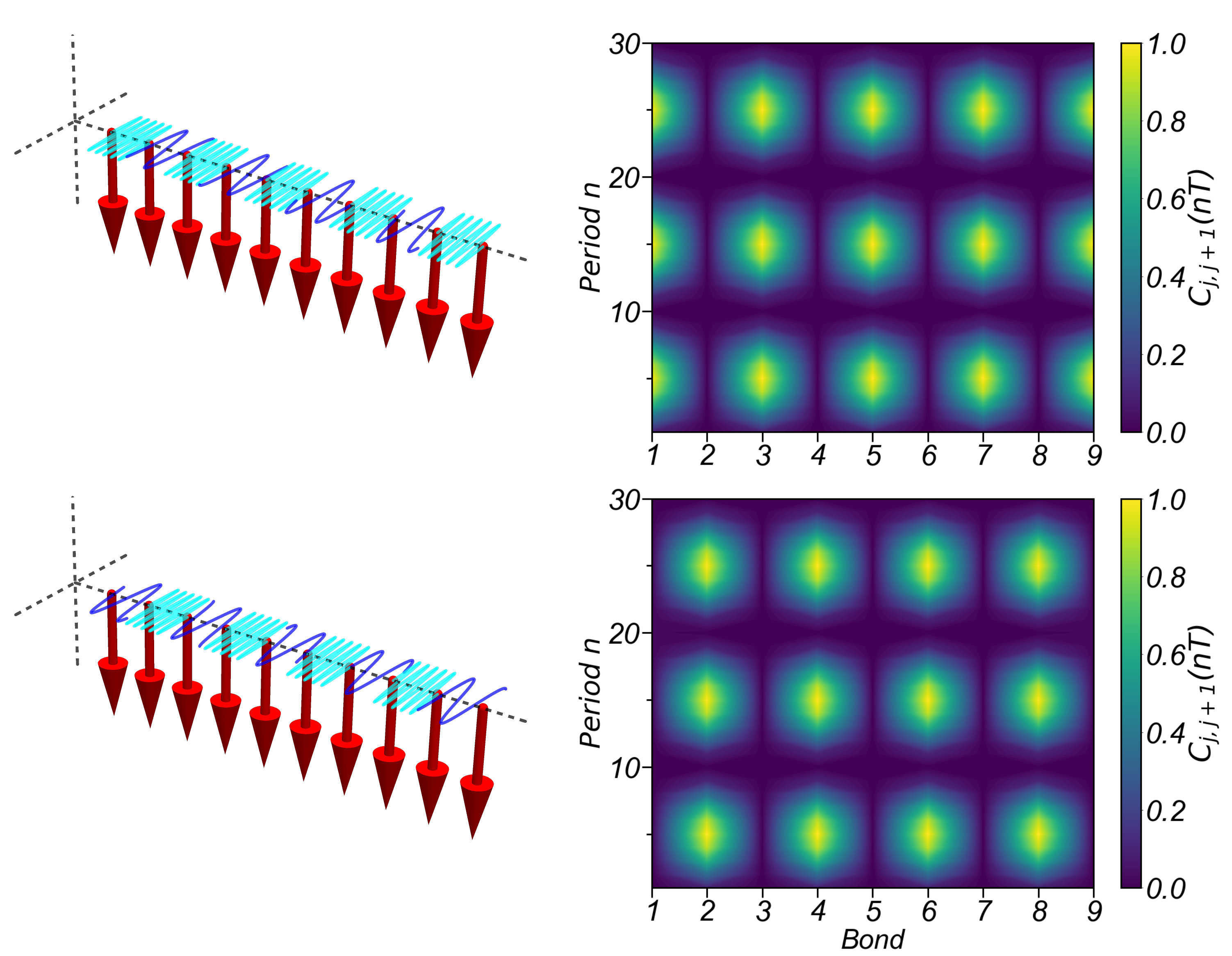}
\caption{Stroboscopic dynamics of nearest-neighbor correlation functions $C_{j,j+1}(nT)$, see the main text. The upper panel shows a lattice where the odd spin exchanges are driven with frequency $\Omega_1$ whereas even spin exchange with frequency $\Omega_0$. The lower panel shows a lattice where the odd spin exchanges are driven with frequency $\Omega_0$ whereas even spin exchange with frequency $\Omega_1$. We consider a spins lattice of size $L=10$ described by the Hamiltonians (\ref{MB}) and (\ref{MB1}) for the upper and lower panels, respectively. The spin system was initialized in the product state $\ket{\psi_0}=\bigotimes_{j=1}^{L}\ket{\downarrow_j}$, and we use the parameter $J_0=0.1g$ and time step $\delta t=0.001g^{-1}$.}
\label{Fig5}
\end{figure}

From Hamiltonian (\ref{MBHI}), we calculate the zero-order term of the Floquet Hamiltonian (\ref{HF01}) 
\begin{align}
\hat{H}^{(0)}_{F,1}=\frac{\hbar J_0}{2}\sum^{L-1}_{j~\rm odd}\big(\hat{\sigma}^{+}_j\hat{\sigma}^{+}_{j+1} + \hat{\sigma}^{-}_j\hat{\sigma}^{-}_{j+1}\big).
\label{HFMB}  
\end{align}
Notice that the Hamiltonian (\ref{HFMB}) involves only interaction terms between spins connected by odd spin exchanges. Consequently, the many-body lattice with inhomogeneous modulated spin exchanges leads to spatial localization of correlated spin pairs. The latter is reflected in the dynamics of non-local observables, such as nearest-neighbor correlation functions. The upper panel in figure \ref{Fig5} shows the stroboscopic dynamics of $C_{j,j+1}(nT)$ for a lattice of size $L=10$ initialized in the product state $\ket{\psi_0}=\bigotimes_{j=1}^{L}\ket{\downarrow_j}$. Here, it is clear that spins connected by odd spin exchanges exhibit correlation, whereas those connected by even spin exchanges lack correlation.   

On the other hand, we interchange the driving protocol in the spins system such that the odd spin exchange is driven with frequency $\Omega_0$ and even spin exchange with frequency $\Omega_1$. The Hamiltonian reads
\begin{align}
\hat{H}(t)&=\hbar g\sum^L_{j}\hat{\sigma}_j^z +  \hbar J_0\cos{(\Omega_0 t)}\sum^{L-1}_{j~\rm odd}\hat{\sigma}^{x}_j\hat{\sigma}^{x}_{j+1} + \hbar J_0\cos{(\Omega_1 t)}\sum^{L-1}_{j~\rm even}\hat{\sigma}^{x}_j\hat{\sigma}^{x}_{j+1}.
\label{MB1} 
\end{align}
If we calculate the zero-order term of the Floquet Hamiltonian, we will obtain
\begin{align}
\hat{H}^{(0)}_{F,2}=\frac{\hbar J_0}{2}\sum^{L-1}_{j~\rm even}\big(\hat{\sigma}^{+}_j\hat{\sigma}^{+}_{j+1} + \hat{\sigma}^{-}_j\hat{\sigma}^{-}_{j+1}\big).
\label{HFMB1}  
\end{align}
Here, it is clear that the system dynamics will be dominated by the spin fluctuations between spins connected by even spin exchange; see the lower panel in figure \ref{Fig5}. These dynamic features are maintained throughout the system's evolution independent of the lattice size $L$. We stress that the behavior between the upper and lower panels in Fig \ref{Fig5} are quite different, and this is so because Hamiltonians (\ref{HFMB}) and (\ref{HFMB1}) are not related by a symmetry transformation when $L$ is even. In contrast, when $L$ is odd, a symmetry with respect to the central spin is present, as we have previously mentioned in the case of the three-spin lattice.

In the previous results where we used the Magnus expansion, we proved the emergence of the spatial localization of correlated spin pairs in the odd or even spin exchanges depending on the two-tone driving protocol. In the next section, we will use these two different two-tone Floquet protocols leading to one-period evolution operators $\hat{U}_1(T)$ and $\hat{U}_2(T)$, each associated with $H^{(0)}_{F,1}$ and $H^{(0)}_{F,2}$, respectively. Those one-period operators will be consecutively applied to the initial state to control the spatial and temporal localization of correlated spin pairs along the lattice.

\section{Steering spin pair fluctuations}
\label{sec:4}

Here, we propose a novel Floquet engineering protocol for steering spin pair fluctuations in the one-dimensional spin-lattice. We will use two different evolution operators $\hat{U}_1(T)$ and $\hat{U}_2(T)$, each associated with the Hamiltonians (\ref{MB}) and (\ref{MB1}), respectively. These operators will be consecutively applied to the initial product state. Operators $\hat{U}_1(T)$ and $\hat{U}_2(T)$ are numerically computed using exact diagonalization.

We define our Floquet protocol in the following way.  Assume that at $t = 0$, the spins system is initialized in the product state $\ket{\psi_0}=\bigotimes_{j=1}^{L}\ket{\downarrow_j}$. Then, we apply the evolution operator $\hat{U}_{1}(T)$ over $m$ periods, and consecutively apply $\hat{U}_{2}(T)$ over $m$ periods, thus completing a $2mT$ evolution time. For a better understanding, we will define the operator $\tilde{U}_j(mT)=\hat{U}^{m}_{j}(T)$ with $j=1,2$. Therefore, our Floquet protocol consists of the consecutive application of the operators $\tilde{U}_1(mT)$ and $\tilde{U}_2(mT)$, which will lead to the control of both spatial and temporal localization of correlated spin pairs The number of periods $m$ is chosen as the ratio $\Omega_0/(2J_0)$, which, according to the values of the parameters used in our numerical simulations $m=10$. It is important to emphasize that this Floquet protocol requires a fine-tuning of the frequencies $g$ and $J_0$, as the ratio $m$ must be either an integer or very close to one. Any deviation from this condition would lead to the accumulation of phase shifts in the many-body wavefunction, resulting in errors that can propagate easily. 

\begin{figure}[t]
\centering
\includegraphics[scale=0.2]{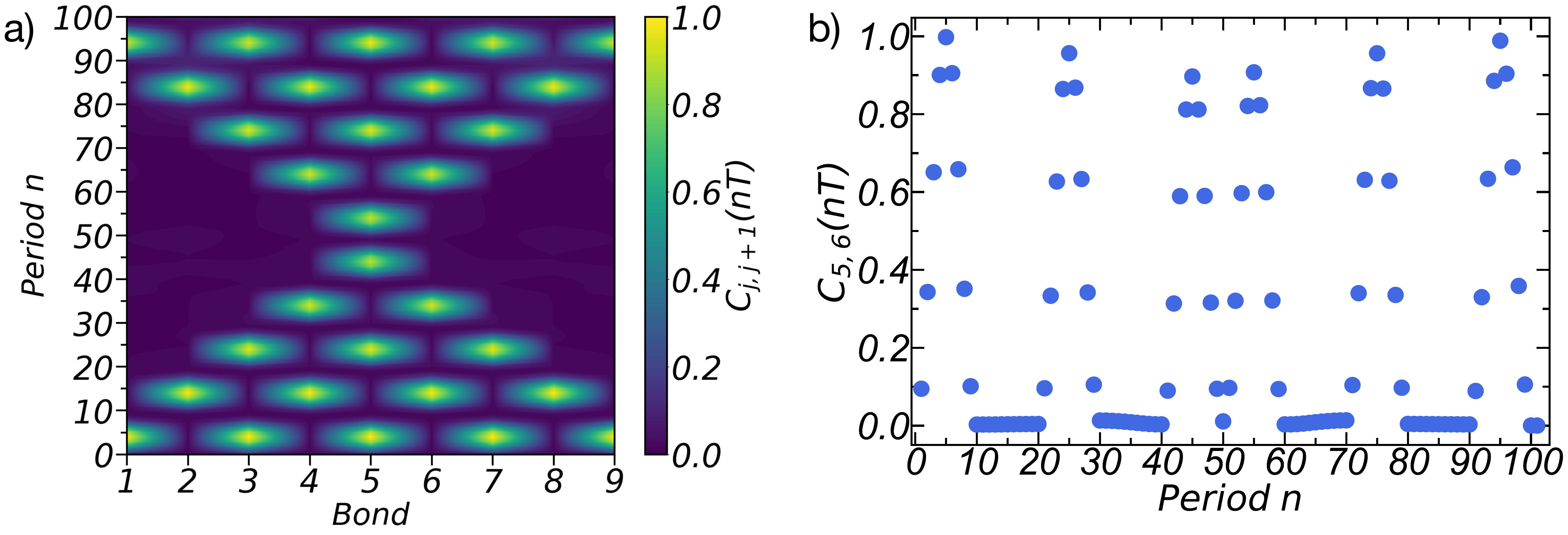}
\caption{a) Nearest-neighbour correlation functions $C_{j,j+1}(nT)$ as a function of stroboscopic time following the Floquet protocol shown in Eq.~(\ref{qTree}), for a spins lattice of size $L=10$. b) Correlation function $C_{5,6}(nT)$ between central spins, whose maximun value reaches $0.9977$ at $t=5T$. As in previous numerical calculations, we use the parameter $J_0=0.1g$ and time step $\delta t=0.001g^{-1}$. The operators $\hat{U}_1(T)$ and $\hat{U}_2(T)$ coming from the two-tone Floquet engineering were computed using exact diagonalization.}
\label{Fig6}
\end{figure}

Now we analyze the case of $L=10$ sites. Figure~\ref{Fig6}(a) shows nearest-neighbor correlation functions $C_{j,j+1}(nT)$ as a function of stroboscopic time following the Floquet protocol. As we consecutively apply the evolution operators $\tilde{U}_{1}(10T)$ and $\tilde{U}_{2}(10T)$ on the initial product state, local spin-correlated pairs emerge in the spins lattice. Initially, these local spin-correlated pairs generate from the edge until they surround the two spins in the center of the lattice, whose correlation function reaches a maximum of $0.9977$ at $t=5T$, see figure \ref{Fig6}(b). In figure~\ref{Fig6}(a), we can see the generation of spatial and temporal localization of correlated spin pairs via dynamically breaking quasi-maximally correlated spin pairs from the edges towards the center of the lattice, up to the time $t=50T$. The quantum dynamic is reversed, and the Floquet engineering starts generating local spin-correlated pairs. The state of the whole sequence shown in figure \ref{Fig6} reads
\begin{align}
\ket{\psi(100T)}=\tilde{U}_{1}\tilde{U}_{2}\tilde{U}_{1}\tilde{U}_{2}\tilde{U}_{1}\tilde{U}_{1}\tilde{U}_{2}\tilde{U}_{1}\tilde{U}_{2}\tilde{U}_{1}\ket{\psi_0}.
\label{qTree}
\end{align}
At the time $t=100T$, the spins system returns to a state close to the initial product state, where the nearest-neighbor spins are negligibly correlated.

\section{Conclusions}
\label{Cl}

In summary, we have reported the steering of spin fluctuations in a one-dimensional spin-1/2 lattice based on a two-tone Floquet engineering. We consider a one-dimensional spin-1/2 lattice with periodically modulated spin exchanges using parametric resonances $\Omega_0$ and $\Omega_1=2\Omega_0$ acting upon consecutive spin exchanges. This two-tone Floquet engineering leads to two critical mechanisms. Firstly, it leads to a subharmonic response in local observables, thus spontaneously breaking the discrete-time translational symmetry. Secondly, it leads to spin pair fluctuations, characterized by nearest-neighbor spins oscillating and reaching quasi-maximally correlated states. The emergence of spin pair fluctuations has a direct application in Floquet protocols, proving the control of nonequilibrium dynamics of interacting many-body systems, producing spatial and temporal localization of correlated spin pairs along the spins lattice in a reversible manner. The latter is reached using two different two-tone Floquet protocols leading to one-period evolution operators $U_1(T)$ and $U_2(T)$, which are consecutively applied to the spins system.  Our findings provide novel Floquet protocols using two simultaneous parametric resonances in the many-body system. These protocols allow us to control spin fluctuations and distribute them in a heterogeneous spin chain, which may open new routes for distinct nonequilibrium states of matter and the conduction of quasiparticles in quantum materials.

\section*{Acknowledgments}
We acknowledge the support from Dicyt USACH under grant 5392304RH-ACDicyt, the support from OpenSuperQ+100 (Grant No. 101113946) of the EU Flagship on Quantum Technologies, the support from Fondo Nacional de Investigaciones Cient\'ificas y Tecnol\'ogicas (FONDECYT, Chile) under grants 1211902 and Centro de Nanociencia y Nanotecnolog\'ia CEDENNA, Financiamiento Basal Puente para Centros Cient\'ificos y Tecnol\'ogicos de Excelencia AFB220001.\\

\bibliographystyle{iopart-num}
\bibliography{Mybib}
\end{document}